\documentclass[superscriptaddress]{revtex4}

\usepackage{graphicx}

\usepackage{bm}
\usepackage{graphicx}
\usepackage{hyperref}
\usepackage{color}

\begin{document}

\title{Yielding and large deviations in micellar gels: a model}

\author{Saroj Kumar Nandi}
\email{snandi@physics.iisc.ernet.in}
\affiliation{Centre for Condensed Matter Theory, Department of Physics, Indian Institute of Science, Bangalore 560012, India}

\author{Bulbul Chakraborty}
\email{bulbul@brandeis.edu}
\affiliation{Martin Fisher School of Physics, Brandeis University, Mail Stop 057, Waltham, Massachusetts 02454-9110, USA}

\author{A. K. Sood}
\email{asood@physics.iisc.ernet.in}
\affiliation{Department of Physics, Indian Institute of Science, Bangalore 560012, India} 

\author{Sriram Ramaswamy}
\email{sriram@tifrh.res.in}
\altaffiliation{On leave at TIFR Centre for Interdisciplinary Sciences, 21 Brundavan Colony, Narsingi, Hyderabad 500 075, India}
\affiliation{Centre for Condensed Matter Theory, Department of Physics, Indian Institute of Science, Bangalore 560012, India}

\begin{abstract} We present a simple model to account for the rheological
behavior observed in recent experiments on micellar gels. The
model combines attachment-detachment kinetics with stretching due to shear, and
shows well-defined jammed and flowing states. The large deviation function
(LDF) for the coarse-grained velocity becomes increasingly non-quadratic as the applied force $F$ is increased, in a range near the yield threshold. The power
fluctuations are found to obey a steady-state fluctuation relation (FR) at
small $F$. However, the FR is violated when $F$ is near the transition from the flowing to the jammed state although the LDF still exists; the antisymmetric part of the LDF is found to be
nonlinear in its argument. Our approach suggests that large fluctuations and
motion in a direction opposite to an imposed force are likely to occur in a
wider class of systems near yielding.  \end{abstract}

\maketitle


Fluctuation relations (FR) as originally formulated
\cite{evans93,gallavotti95a,gallavotti95b,jarzynski97,crooks98,bochkov77,bochkov81} are
exact statements connecting the relative probabilities of observing the
production and the consumption of entropy at a given rate in a driven
\textit{thermal} system, and can be expressed as symmetry properties of the
large-deviation function of the entropy production rate. An observable
$X_{\tau}$, for example, the average power delivered over a time interval $\tau$ is said to have the
large-deviation property if its probability density decreases exponentially for
large $\tau$, $\mbox{Prob}(X_{\tau\to\infty} = a) \sim \exp[-\tau
\mathcal{W}(a)]$. The decay rate $\mathcal{W}$ is called the large-deviation
function (LDF), and its behaviour for large $a$ encodes information about the
statistics of extremes of the underlying random process \cite{touchette09,oono89}. The
steady-state fluctuation relation \cite{gallavotti95a,gallavotti95b} states that
$\mathcal{W}(a) - \mathcal{W}(-a) \propto a$.  

In this paper we study numerically in some detail the large-deviation behaviour
of a model for a
macroscopic degree of freedom driven through a medium of dynamic attachment
points. The model is motivated by precision creep rheometry studies 
\cite{sayantan08} of a micellar gel at controlled stresses below its nominal
yield point, that revealed that the small positive mean shear-rate, i.e., in the
direction favoured by the imposed stress, was composed of a highly irregular
time-series of positive and negative shearing events. Moreover the shear-rate
fluctuations, which at constant stress are also power fluctuations, obeyed
\cite{sayantan08} a fluctuation relation of the Gallavotti-Cohen type
\cite{gallavotti95a,gallavotti95b}. The measurements reported were made on a
macroscopic degree of freedom, the angular position of the rheometer plate,
which could not be influenced perceptibly by thermal noise. Fluctuations in this
system must be a consequence of the imposed drive. The probability distribution
of the angular-velocity fluctuations was found to be strongly non-Gaussian at
large imposed stress. An effective temperature extracted from a comparison to
the Gallavotti-Cohen relation was found to increase with the applied stress.

Our theoretical model reproduces the findings of the experiment
but also finds departures from the fluctuation relation in the strict
sense, in a certain parameter range, despite the existence of LDF. That is, it
finds that the antisymmetric part of the LDF departs from linearity in its
argument. Moreover, the model we present should apply to a wider class
physical problems involving yield or escape in the presence of fluctuations. 
We shall return to these points at the end of the paper.

Consider an external force acting on some coordinate of a Hamiltonian system obeying the chaotic hypothesis \cite{gallavotti95a,gallavotti95b} with inverse temperature $\beta$, and let the random variable $w_t$ be the instantaneous rate of doing work in a given realization of the dynamics. Define
\begin{equation}
W_\tau=\frac{1}{\tau}\int_0^\tau w_t dt 
\end{equation}
to be the rate of doing work, binned or averaged over a time scale
$\tau$. The steady-state fluctuation relation (FR)
\cite{wang05,evans02} tells us 
\begin{equation}
\label{SSFR}
\frac{1}{\tau}\ln\frac{P(W_\tau=w)}{P(W\tau=-w)} \asymp {\beta w}, \,\, {\tau\to\infty}
\end{equation}
where $P(W_\tau)$ is the probability distribution function (PDF) of $W_\tau$,
and $\asymp$ denotes asymptotic equality. Underlying (\ref{SSFR}) is a more general relation, 
the existence of the large deviation function (LDF) \cite{touchette09}. The principle of large 
deviation consists in the existence of the limit 
\begin{equation}
\label{LDP}
\lim_{\tau\to\infty}-\frac{1}{\tau}\ln P(W_\tau=w)=\mathcal{W}(w),
\end{equation}
where the limiting quantity $\mathcal{W}$ is called the rate function or
large-deviation function (LDF). 
Eq. \ref{SSFR} is then a statement about a symmetry property of the LDF, \textit{viz.},
that its antisymmetric part is linear in its argument. Experiments \cite{douarche06,nitin11,gomez11,feitosa04}, simulations \cite{zamponi05,gradenigo12} and theoretical calculations \cite{sanjib11,vanzonpre03,debarati10,debarati11,ciliberto10} confirm the existence of relations of the FR type in a wide class of systems \cite{seifert12}. By construction \cite{gaspard04} the minimum value of the true $\mathcal{W}$ has to be zero. Experiments and simulations work at finite $\tau$, which inevitably leads to an offset such that $\mathcal{W}$ has a positive minimum. Put another way, implementing Eq. (\ref{LDP}) in practice yields $\mathcal{W}$ upto a positive additive shift which decreases with increasing $\tau$, as can be seen in Figs. \ref{extF20_N200}-\ref{extF86_N200}.
Note that some of
these systems are not obviously characterized by a thermodynamic temperature.
In such cases the existence of a relation like (\ref{SSFR}) offers one way
of defining an effective temperature \cite{cugliandolo11}.    

The exact linearity in $W_\tau$ as required by (\ref{SSFR}) is a strong restriction. Naturally there are situations
\cite{farago02,vanzon03,farago04,harris05,touchette07,baule09,gomezsolano10,sanjibpre,sayantan12,chechkin09} where it does not hold. 
In the example of \cite{vanzon03,touchette07}, the large-deviation property itself does
not hold. However, the FR may be violated in a variety of situations even if the large-deviation function exists. For example, in \cite{farago02,farago04,sanjib11} an external power is injected into the system and when this power injection exceeds a certain value, even though the LDF exists, FR is not obeyed anymore. 
A similar scenario occurs in the experiments of \cite{gomezsolano10} where FR is violated as the system is driven out-of-equilibrium beyond a certain limit. However, the existence of LDF in this regime has not been analysed in \cite{gomezsolano10}. Ref. \cite{baule09} analyses two models: first, a dragged particle is subjected to an external Poissonian shot noise (PSN) and second, it is subjected to a Gaussian thermal noise in addition to PSN. In both the cases, the LDF exists but the steady state fluctuation relation is violated. The results in this work also show a similar trend, in our simple model, the conventional FR fails to hold in some regions of parameter space even though the LDF exists.  
The model has the additional virtue of being constructed to model a physical situation,
rather than as a mathematical counterexample, and will therefore be of wide interest. 

Here are the main results of this work: (1) the model shows a sharp crossover from a creeping jammed state to steady flow (Fig. \ref{flow_curve}); (2) deep in the jammed state, the LDF of velocity fluctuations is quadratic and they obey the FR (Fig. \ref{extF20_N200}); (3) near the threshold to free flow, the LDF becomes non-quadratic, but the velocity fluctuations still obey the FR (Fig. \ref{extF86_N200}); all these findings are in conformity with experiments \cite{sayantan08}. (4) Just below the threshold, the LDF becomes non-quadratic and the velocity fluctuations don't obey FR as shown in Fig. \ref{extF96_N200}).
(5) If we keep the number of attachment points fixed, the effective temperature ($T_{eff}$) decreases with increasing applied force ($F)$ in contrast to what is found in the experiment. To produce the correct $T_{eff}$ vs $F$ trend, we must allow the number of attachment points to increase with increasing $F$ (Fig. \ref{effectiveT}).

The paper is organised as follows: in Sec \ref{model} we present our model and
we discuss the mean-field version of the model in Sec. \ref{meanfield}. We present details of our exploration of parameter space and the results in Sec. \ref{results} and conclude the paper with a
discussion in Sec. \ref{discussion}.

\section{The model}
\label{model}
The micellar gel sample in the experiment is taken in a rheometer with a cone-plate geometry to ensure uniform strain rate through-out the sample, and the upper plate is rotated with a constant torque while keeping the lower plate fixed. The detailed geometry of the rheometer, however, is not important for the observed findings. 
We model the micellar gel medium as a collection of springs that are stretched by the applied torque when they are attached to the plates, and can detach when stretched by a high enough force. We do not associate the ``springs'' with individual molecules or micelles but rather with adhering, deformable domains in the material, whose size we do not know.
We assume the springs always remain attached to the stationary plate. 
\begin{figure}[h]
\begin{center}
\includegraphics[height=5cm]{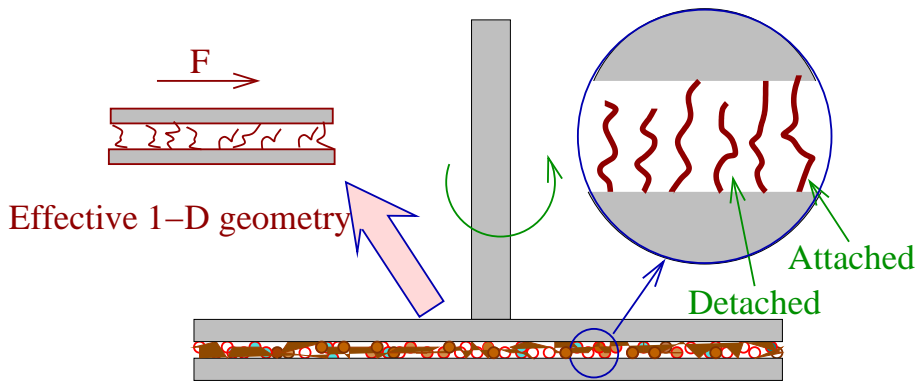}
\end{center}
\caption{A schematic illustration of the model. The ``springs'' (see text) can get attached to or detached from the upper plate. The experimental geometry can be thought of as effectively one-dimensional with the force being applied in a particular direction.}
\label{rheology_schematic}
\end{figure}

We use an effective one-dimensional description in which $X(t)$ is the total (angular) displacement of the upper plate as in Fig. \ref{rheology_schematic} and let $V(t)=dX(t)/dt$ be its instantaneous velocity. 
Let us consider $f_i(t)=k_ix_i$ be the force on the $i$th spring at time $t$ where $k_i$ and $x_i$ are respectively the spring constant and the extension of the spring. Then the spring will pull the plate backwards only if it is attached with the plate. Thus, we can write down the equation of motion for the upper plate as
\begin{equation}
\label{modeleq1}
M\frac{dV(t)}{dt}+BV(t)=F-\sum_i s_if_i(t),
\end{equation}
where $M$ is the mass of the plate, $B$ a viscous damping coefficient, $F$ the force (actually torque) on the rheometer plates and $s_i$ is a two state variable which can take on values $0$ or 1. If $s_i=1$, then the $i$th spring is attached with the plate and it is detached otherwise.
Note that we have not specified the number $N$ of ``springs'' in the model. It is not clear how to do this in the absence of a detailed microscopic theory. It is entirely possible that $N$ depends on the imposed torque, or even that it is determined dynamically. We will assume it is a parameter of the system, and show the behavior in the $N-F$ plane.

Now, if the spring is attached to the plate, it will stretch with the
velocity of the plate; if it is detached it will relax. Thus 
\begin{equation}
\label{force_relax}
\frac{dx_i(t)}{dt}=-(1-s_i)\gamma k_ix_i(t)+s_iV(t)
\end{equation}
where $\gamma$ is a kinetic coefficient. Thus,
\begin{equation}
\label{forcerelax2}
\frac{df_i(t)}{dt}=-(1-s_i)\gamma k_i f_i(t)+s_ik_iV(t).
\end{equation}
Assume for simplicity that all $k_i$'s are equal, define $\gamma k_i\equiv 1/\tau$, redefine $k_iV\to V$ in Eq. (\ref{forcerelax2}) and $B/k_i\to B$ in Eq. (\ref{modeleq1}) and then we set $B$ to unity. Ignoring inertia, we then obtain equations for $V$ and $f_i(t)$: 
\begin{eqnarray}
\label{model1}
V(t)=F-\sum_i s_if_i(t) \\
\label{model2}
\frac{df_i(t)}{dt}=-(1-s_i) f_i(t)/\tau+s_iV(t).
\end{eqnarray}
One could imagine more complicated modes of relaxation, for example the springs
can partly redistribute forces among themselves. But Eq. (\ref{model2}) is the
simplest possible model that contains the dominant mechanism. We will discuss
the effect of a diffusive term later.

The state variables $s_i$ are assumed to follow a stochastic dynamics. Let $P_i(t)\equiv$Prob($s_i=1,t$) be the probability that the $i$th spring is attached at time $t$. Then 
\begin{equation}
\label{modeleq3}
\frac{dP_i(t)}{dt}=-W_DP_i(t)+W_A(1-P_i(t)),
\end{equation}
where $W_A$ and $W_D$ are the attachment and detachment rates, the most
important input parameters of the model. Both $W_A$ and $W_D$ depend on $f_i$. It is possible to engineer these rates
to reproduce different behaviours by the model. The feature that is essential
to get negative fluctuations is that a spring with a large force on it gets
reattached and pulls the plate in the opposite direction.
For a suitable choice of parameter values, as
will be shown below, the model shows a jammed-flowing transition. This is not a
true phase transition, but a strong crossover from slow creep to free flow.

\section{The mean field calculation}
\label{meanfield}
In the mean field approximation, we replace the forces and the $s_i$ by their average values which we take to be the same for all $i$: $\langle f_i\rangle=f$ and $\langle s_i\rangle=s$ so that
\begin{equation}
\label{mfv}
\langle V\rangle=F-\sum_i\langle f_i s_i \rangle \simeq F-\sum_i\langle f_i\rangle\langle s_i \rangle=F-Nsf,
\end{equation}
where $N$ is the total number of springs. In the steady state, (\ref{model2}) will yield
\begin{equation}
(1-s)f/\tau=s\langle V\rangle \Rightarrow f= \tau_1 s\langle V\rangle/(1-s).
\end{equation}
Using the above relation in Eq. (\ref{mfv}), we find the velocity 
\begin{equation}
\langle V\rangle=\frac{(1-s)F}{(1-s)+Ns^2\tau}.
\end{equation}
In two extreme limits, if $s=1$, $\langle V\rangle=0$ and if $s=0$, $\langle V\rangle=F$. In the steady state, $dP_i(t)/dt = 0$ so that
\begin{equation}
P_i=\frac{W_A}{W_A+W_D}.
\end{equation}
Therefore, the steady state value of $s$, within the mean-field approximation, will be
\begin{equation}
s=\langle s_i\rangle = \sum_{s_i=0,1} s_iP_i= P_i=\frac{W_A}{W_A+W_D}.
\end{equation}

The mean force on each spring becomes
\begin{equation}
f=\frac{\tau sF}{(1-s) + Ns^2\tau}.
\end{equation}

\begin{figure}
\begin{center}
\includegraphics[height=5cm]{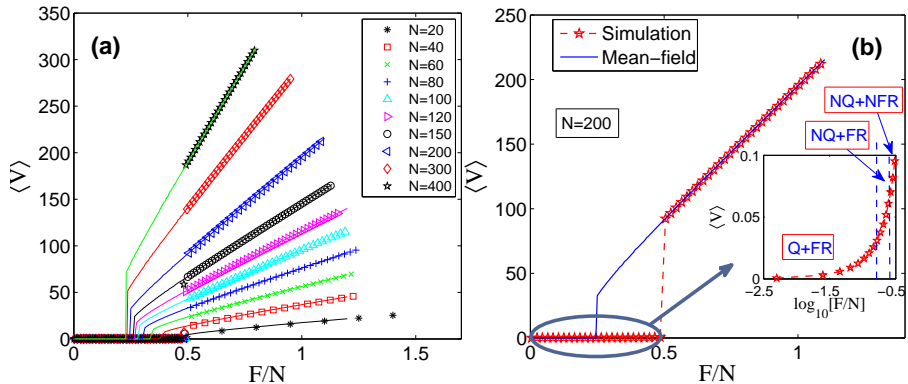}
\end{center}
\caption{(a) The flow curves obtained from the simulation for various number $N$ of springs with a particular set of parameter values (see text), the lines are the corresponding mean-field solution. (b) The flow curve is shown for $N=200$ for clarity. The mean-field solution underestimates the threshold. The regions are roughly marked based on whether the LDF is quadratic (Q) or non-quadratic (NQ) and whether the fluctuations obey the conventional FR or not.}
\label{flow_curve}
\end{figure}

The predictions of mean-field version of the model with the particular set of input parameter values as used for the simulation are shown with solid lines in Fig. \ref{flow_curve}.

\section{Details of the simulation and the results}
\label{results}
We simulate the equations (\ref{model1})-(\ref{model2}) and the stochastic dynamics of the $s_i$ corresponding to (\ref{modeleq3}) through the kinetic Monte-Carlo (KMC) method \cite{kmc} to obtain the behaviour of our model. The advantage of KMC over the conventional Monte-Carlo method is that the time scale of the dynamics is entirely determined by the various rates of the problem. 

We see from Eq. (\ref{mfv}) that $\langle V\rangle$ depends on three time scales. To simplify
the discussion, we fix $\tau$ and $W_A$ and take $W_D/W_A$ to have an activated
form. We recall that the mechanism of having a negative velocity (in the
direction opposite to $F$) events is that a spring with a large force on it
gets reattached to the plate before it has completely relaxed its force. This
can happen if the force relaxation is much slower than the
attachment-detachment kinetics of the springs. Thus, to have a large number of
negative events, we must have $\tau W_A\gg1$. To ensure this, we chose the
parameters as follows: $\tau=2.5$, $W_A=100$ and $W_D=W_Ae^{\alpha(f_i-f_0)}$,
with $\alpha=2.0$ and $f_0=1.0$. The springs can get attached with the plate at
a constant rate irrespective of the force on it. But if it is already attached
to the plate, it is more probable to get detached as the force on the spring
increases. The relaxation time of the attachment-detachment kinetics of the springs with these parameter values and $N\sim 100$ is of the order of $10^{-4}$. We have introduced the parameter $\alpha$ to obtain a reasonably
sharp  transition from jammed to flowing state, at a force whose value is
controlled by $f_0$. Restrictig the form of $W_D$ allows us to worry about one
less parameter of the model.

The activated nature of the attachment-detachment is reminiscent of rheology
models with traps such as the SGR \cite{sollich97,sollich98,fielding00}. An essential difference between our model and the trap models \cite{sollich97,sollich98} is that the springs in our model
retain the forces on them even after getting detached from the plate whereas
the strain on a spring in the trap model becomes zero after it comes out of a
trap. The simplest trap models can not show negative velocity events.

The model shows a well-defined jammed to flowing transition that becomes sharper as we increase the number of springs $N$ as is shown in Fig. \ref{flow_curve} where the symbols are the simulation values and the corresponding curves are the mean-field solution. We see that the mean-field solution underestimates the threshold since mean-field overlooks the fluctuations and noise makes depinning easier. In Fig. \ref{flow_curve}(b) we show the flow curve for $N=200$ and the inset shows the behaviour in the jammed state in a semi-log plot where it is evident that even in the jammed state, the velocity is actually non-zero. The mean-field solution overestimates the threshold but agrees well with simulation results away from the transition.

\begin{figure}
\begin{center}
\includegraphics[width=12cm]{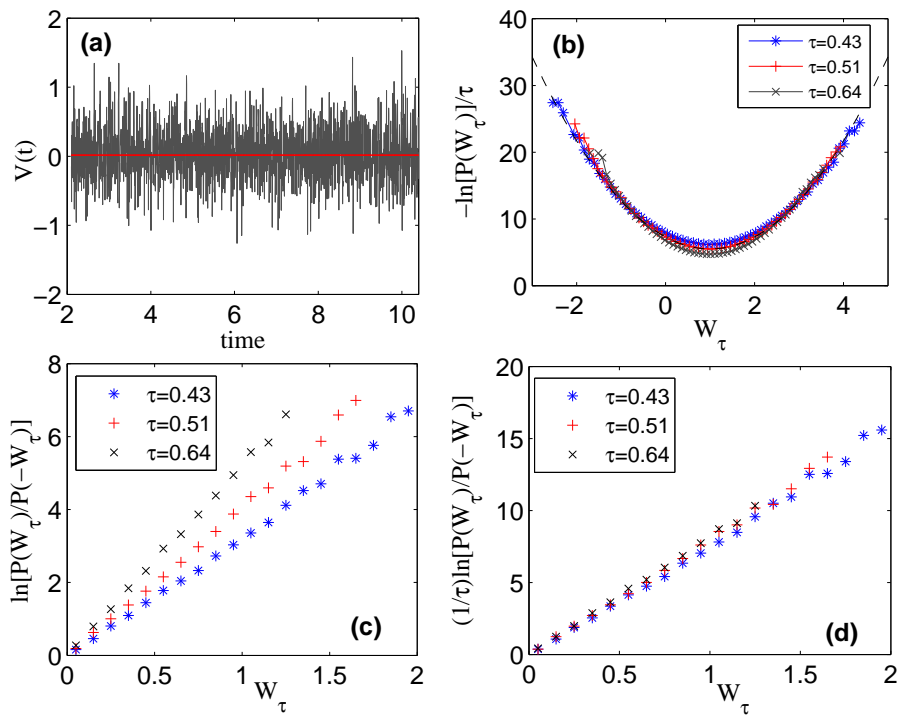}
\end{center}
\caption{The behaviour of the model for the particular set of parameter values as specified in the text with $N=200$ and $F=20.0$ (a) The instantaneous velocity as a function of time, the thick line denotes the average velocity. There is a significant number of negative velocity events. (b) The large-deviation function (LDF) for the scaled coarse-grained velocity or work fluctuation (both are same since the applied force is constant) $W_\tau$. The quadratic function is shown by the dotted curve. We see that the LDF is quadratic. (c) $\ln[P(W_\tau)/P(-W_\tau)]$ vs $W_\tau$ for various $\tau$. (d) $(1/\tau)\ln[P(W_\tau)/P(-W_\tau)]$ vs $W_\tau$ for various $\tau$ collapse to a master curve that is straight line signifies that the velocity fluctuation obeys fluctuation relation.}
\label{extF20_N200}
\end{figure}

\begin{figure}
\begin{center}
\includegraphics[width=15cm]{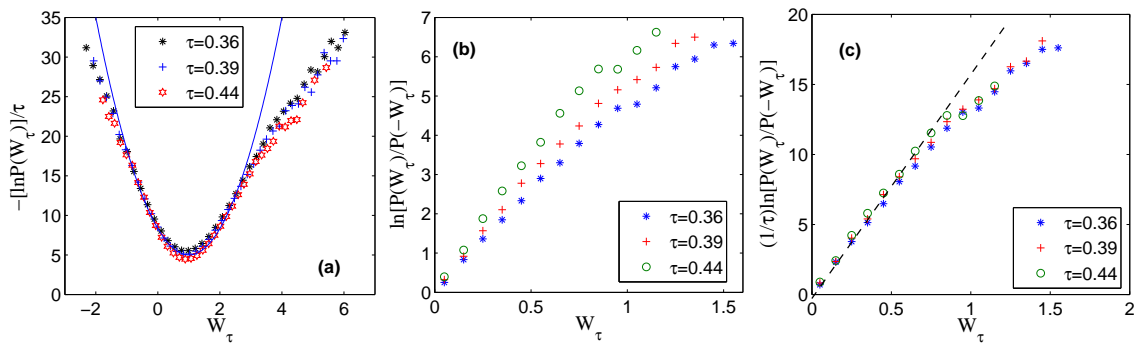}
\end{center}
\caption{The number of springs is $N=200$ and $F=96.0$ which is very close to the threshold value. (a) The large-deviation function exists for the velocity fluctuation $W_\tau$ and it is non-quadratic. The quadratic function is also shown with the solid line. (b) $\ln[P(W_\tau)/P(-W_\tau)]$ vs $W_\tau$ for various $\tau$ as shown in the figure. The curves deviate from the expected straight line if FR was obeyed by $W_\tau$. (c) When we scale the various curves in (b) by $\tau$, they show excellent data collapse, but the collapsed data deviates significantly from a straight line implying the violation of the conventional FR. The straight line in the figure is just a guide to the eye.}
\label{extF96_N200}
\end{figure}

In the jammed state, there is a significant number of negative velocity events, however, there are none once the system goes to the flowing state. To understand the fluctuations in the jammed state and test the regime of validity of the FR, we keep $N=200$ fixed, and find a threshold force $F=97.0$. We take a value of the external force $F=20.0$ deep in the jammed state. From the instantaneous velocity as shown in Fig. \ref{extF20_N200} (a), we see that there is a significant number of negative velocity events. Since the applied force is constant, the statistics of velocity and power fluctuations are the same. We denote the velocity (or power) fluctuations with respect to the mean, averaged over a time interval $\tau$ by $W_\tau$:
\begin{equation}
W_\tau=\frac{1}{\tau}\int_T^{T+\tau}\frac{V(t)}{\langle V\rangle}dt.
\end{equation}
We sample time intervals separated by durations greater than the mean correlation time of the velocity. The LDF for the velocity fluctuation $W_{\tau}$ is shown in Fig. \ref{extF20_N200}(b). The LDF is found to be quadratic. Also, $W_\tau$ obeys the fluctuation relation as is evident from the plot of $\ln[P(W_\tau)/P(-W_\tau)]/\tau$ vs $W_\tau$ for various $\tau$ collapsing to a master curve that is a straight line going through the origin [Fig. \ref{extF20_N200}(d)].

One of the interesting features of the model is that if we are very close to the threshold, even though the large deviation function exists, $W_\tau$ doesn't obey the standard fluctuation relation, as is seen in Fig. \ref{extF96_N200}. 
Here we keep the applied force $F=96.0$ which is very close to the threshold value. In this case, the LDF becomes non-quadratic as is evident from Fig. \ref{extF96_N200}(a). The plot of $\ln[P(W_\tau)/P(-W_\tau)]$ vs $W_\tau$ deviates from a straight line and if we scale $\ln[P(W_\tau)/P(-W_\tau)]$ by $\tau$, even though we obtain data collapse, the master curve is no longer a straight line  as shown in Fig. \ref{extF96_N200}(c). As long as we are very close to the threshold force, similar nonlinear FR curves are obtained even if we change the number of springs $N$. 
Our simple model thus offers an example of a system with substantial negative fluctuations and excellent data collapse consistent with the large-deviation property, but in which the antisymmetric part of the large-deviation function is strongly nonlinear.

\begin{figure}
\begin{center}
\includegraphics[width=15cm]{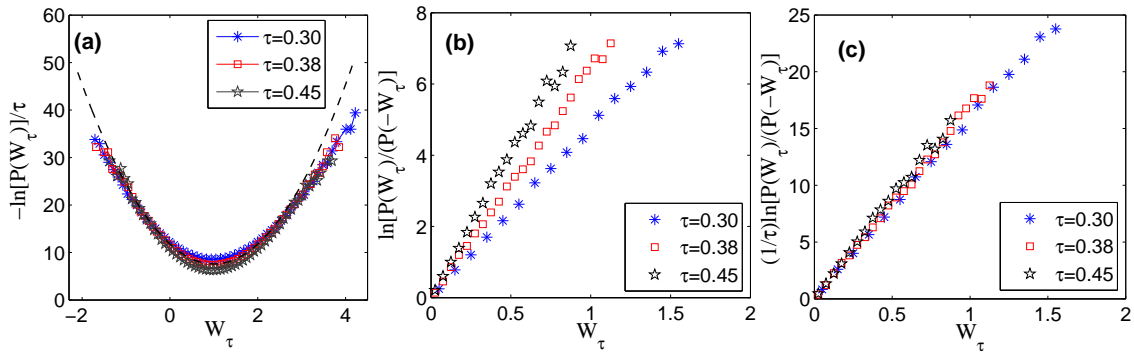}
\end{center}
\caption{The number of springs is $N=200$ and applied force $F=86$ which is close to the threshold value. (a) The large-deviation function deviates significantly from the quadratic function (dotted curve) near its tail. (b) Plot of $\ln[P(W_\tau)/P(-W_\tau)]$ vs $W_\tau$ for various $\tau$ as shown in the figure. (c) When we scale $\ln[P(W_\tau)/P(-W_\tau)]$ with $\tau$, the curves show data collapse and the master curve is a straight line going through the origin ascertaining the validity of FR.}
\label{extF86_N200}
\end{figure}

However, if we move slightly away from the threshold but still within the jammed state, the LDF remains non-quadratic, but the fluctuation relation is obeyed over the entire range of our data (Fig. \ref{extF86_N200}). This shows that the source of deviation from the fluctuation relations is not merely the non-quadratic nature of the LDF, but it is an intrinsic nature of the model and stems from a complex mechanism near the threshold. We do not completely understand the origin of such behaviour, but further work in this direction should elucidate this very interesting phenomenon. 

\begin{figure}
\begin{center}
\includegraphics[height=3cm]{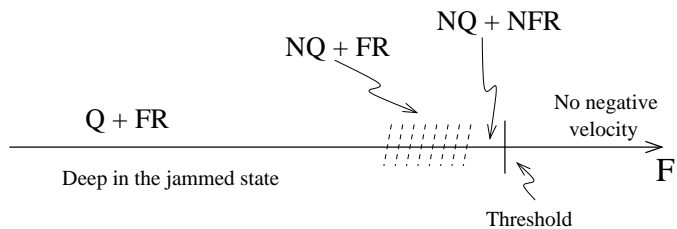}
\end{center}
\caption{For a certain number of springs, the system goes from a jammed creeping state at low external force ($F$) to a free flowing regime as $F$ is increased. Deep in the jammed state, the LDF of the velocity fluctuation is quadratic (Q) and obeys FR. However, as we increase $F$ towards the transition, close to the transition, LDF becomes non-quadratic (NQ) although FR is still obeyed. Very close to the threshold, the LDF is NQ and FR doesn't hold anymore. We do not see any negative velocity events beyond the threshold.}
\label{rheology_phasedia}
\end{figure}
Let us summarize the behaviour of the model in a schematic phase diagram, Fig. \ref{rheology_phasedia}. Deep in the jammed state, the LDF is quadratic (Q) and the velocity fluctuations obey the fluctuation relations (FR). Near the threshold force of the jammed-to-flowing transition, the LDF becomes non-quadratic (NQ) but the velocity fluctuations still obey FR. As we approach the transition, very close to the threshold, the LDF becomes non-quadratic and the velocity fluctuations no longer obey FR (NFR). We note that these features of our model are similar to the statistical properties of entropy-consuming fluctuations in jammed states of laponite suspensions \cite{sayantan12}.

\begin{figure}
\begin{center}
\includegraphics[height=5cm]{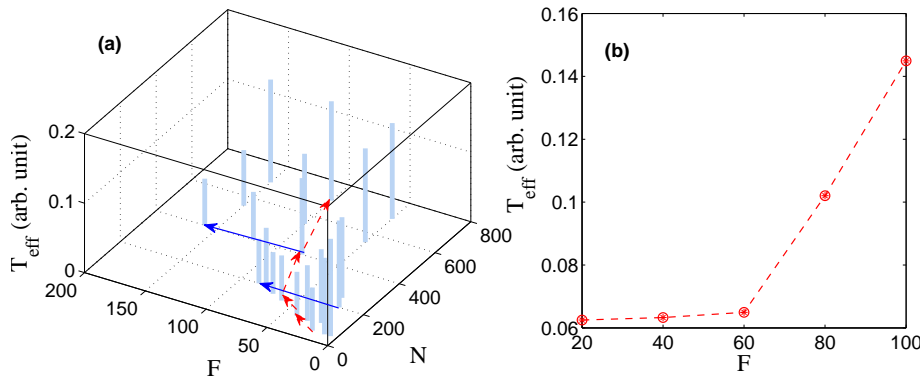}
\end{center}
\caption{
(a) The effective temperature $T_{eff}$ extracted from Eq. (\ref{SSFR}) in arbitrary unit is plotted as a function of $N$ and $F$. If we keep $N$ fixed in the simulation, $T_{eff}$ decreases with increasing $F$ as is found from the two paths shown in the figure by two blue solid arrows corresponding to $N=200$ and $N=400$. However, if we consider a different path, one such possible path is shown by the red dotted arrow, where $N$ increases with increasing $F$, $T_{eff}$ increases on this
path. (b) $T_{eff}$ vs $F$ for the path denoted by the red dotted arrow in (a) is
shown for clarity. The number of filaments corresponding to a particular $F$ is
listed in Table \ref{numbervsF}.
}
\label{effectiveT}
\end{figure}

The slope of the scaled FR plot, analogous to $\beta$ in Eq. (\ref{SSFR}), can be thought of as the inverse of an effective temperature. In the experiment, the effective temperature ($T_{eff}$) increases as $F$ increases \cite{sayantan08}. We have pointed out earlier that there is no reason for the number of attachment points $N$ to remain fixed in the model. In fact, it is more reasonable that $N$ changes with $F$, since large applied force can disentangle or break micelles or disrupt domains giving rise to more independent regions in the dynamics. To see the behaviour of $T_{eff}$ as a function of $F$ within our model, we have plotted $T_{eff}$ as a function of both $F$ and $N$ (Fig. \ref{effectiveT}). Let us first see what happens if we keep $N$ fixed. As we have shown in Fig. \ref{effectiveT}(a) for $N=200$ and $N=400$ (two solid arrows), $T_{eff}$ decreases as $F$ increases with constant $N$. This is in complete contrast to what was found in the experiment \cite{sayantan08}. However, as $N$ increases, $T_{eff}$ increases. Thus, to be consistent with our model, it must be that the system moves on a path on which $N$ changes with $F$. With this in mind, it is possible to identify a path in the 3$d$ space of $(N,F,T_{eff})$ where the model reproduces the correct trend of $T_{eff}$ vs $F$. One such possible path is shown by the dotted arrows in Fig. \ref{effectiveT}(a) and the corresponding $T_{eff}$ vs $F$ behaviours is shown in Fig. \ref{effectiveT}(b) for clarity. We also show in Table \ref{numbervsF} the particular number of attachment points for a particular force $F$ corresponding to this path. We emphasize that this is not the only possible path consistent with the experimental trends, and the particular path that the experiment will follow is going to depend on the microscopic details of the experiment.

\begin{table}
\centering
\begin{tabular}{cccccc}
\hline
F & 20 & 40 & 60 & 80 & 100 \\
N & 50 & 100 & 150 & 400 & 700 \\
\hline
\end{tabular}
\caption{The number of springs $N$ taken at a particular $F$}
\label{numbervsF}
\end{table}

\section{Discussion and conclusion} 
\label{discussion} 
In this work we have presented a simple model to understand a particular set of
experiments where it was found that the velocity (or power, since the applied
force is constant) fluctuations obey the fluctuation relations. The
force-dependent attachment-detachment kinetics of the springs with the plate is
the main mechanism behind the observed negative velocity. When the applied
force is very close to the threshold, the large-deviation function of velocity fluctuations becomes non-quadratic and the strong departures from a conventional fluctuation relation are seen. This is especially interesting
given that the large-deviation property continues to hold, and we obtain data
collapse when we plot $\ln[P(W_\tau)/P(-W_\tau)]/\tau$ as a function of
$W_\tau$, though the master curve is not a straight line. 
A number of theoretical models \cite{farago02,farago04,baule09,sanjib11} of systems driven out of equilibrium by externally imposed noise display such a departure from the conventional FR where the LDF exists. Our model differs from these in that it rationalizes a specific set of experiments on systems near yielding, and relies on the amplification of fluctuations by a deterministic driving force.
We need more experiments and theoretical analysis to understand the
origin of such a phenomenon in our model. 
The observation of a linear dependence of $\ln [P(W_\tau)/P(-W_\tau)]$ over a range of $W_\tau$ is not in itself our major finding, as a function that goes through zero will normally have a linear range. It is of greater significance (a) that the function has appreciable weight at negative arguments, suggesting that the model captures some of the essential physics of the experiment and (b) that we observe good data collapse even when the symmetry function departs from linearity.
A worthwhile future direction will be to sample
the rare events \cite{berryman10,anupam11} to improve statistics in the
tail of the distribution, possibly elucidating the nature of this deviation.

If we keep the number of springs taking part in the dynamics fixed, the observed trend of the effective temperature as a function of applied force is opposite to what was found in the experiment. However, it is more reasonable to vary the number of springs as $F$ changes, since larger applied force may break entanglements, rupture micelles, or disrupt adhering domains. This allows the model to reproduce the correct trend of $T_{eff}$. 

In the model, we have allowed a simple local relaxation mechanism for the springs. One can imagine more complicated modes of relaxation, for example, we can allow the springs to redistribute forces among their neighbors:
\begin{equation}
\frac{df_i(t)}{dt}=(1-s_i)\bigg[-\frac{f_i(t)}{\tau_1}+\frac{-2f_i+f_{i-1}+f_{i+1}}{\tau_2}\bigg]+s_iV(t).
\end{equation}
We find that the presence of such a diffusive term doesn't affect the behaviour of the model much. During the attachment-detachment kinetics, other processes that could play a role include: spatial inhomogeneity and temporal variation of spring stiffness and their modification by local stretching and release, and the interplay of micellar lengths and relaxation time with imposed stresses \cite{nitin12}.

Finally, we expect that fluctuations near yielding in a wider class of systems could show features similar to those discussed here. We have in mind situations such as the dislocation-mediated flow of stressed crystals at non-zero temperature \cite{zippelius80}, the flow of glass through the mechanism proposed by Sausset {\it et al.} \cite{sausset11}, and thermally assisted depinning in general \cite{bustingorry12}. Fluctuations that take a region from the downhill to the uphill side of a pinning barrier, which are clearly more likely to happen near yielding, where effective barriers are small, should give rise to negative-velocity events.

\begin{acknowledgements}
{SKN would like to thank Sayantan Majumdar and Sumilan Banerjee for discussions. BC and SKN thank the TIFR Centre for Interdisciplinary Sciences, Hyderabad, for hospitality. SKN was supported in part by the University Grants Commission and SR by a J.C. Bose Fellowship from the Department of Science and Technology, India. AKS thanks CSIR for support as Bhatnagar Fellowship. BC acknowledges discussions with Peter Sollich, the hospitality of KITP, Santa Barbara where some of this work was done and support from NSF-DMR award 0905880.}
\end{acknowledgements}

\end{document}